\documentclass[referee]{raa}
\usepackage{graphicx,times}
\usepackage{natbib}
\date {}
\begin{document}
\title{Search for pulsations in the LMXB EXO 0748-676}

   \setcounter{page}{1}

   \author{Chetana Jain
      \inst{}
   \and Biswajit Paul
      \inst{}
   }
   \institute{Raman Research Institute, Sadashivnagar, C. V. Raman Avenue,  
             Bangalore 560080, India.\footnote{CJ is currently working at Hans Raj College, University of Delhi, Delhi-110007, India.} {\it chetanajain11@gmail.com}\\
}
\abstract{ We present here results from our search for X-ray pulsations of the neutron star in the low mass X-ray binary EXO 0748-676 at a frequency near a burst-oscillation frequency of 44.7 Hz. Using the observations made with the Proportional Counter Array on board the Rossi X-ray Timing Explorer we did not find any pulsations in the frequency band of 44.4 Hz to 45.0 Hz and obtained a 3$\sigma$ upper limit of 0.47 $\%$ on the pulsed fraction for any possible underlying pulsation in this frequency band. We also discuss the importance of EXO 0748-676, as a promising source for detection of Gravitational Waves.
}
   \authorrunning{C. Jain \& B. Paul}            
   \titlerunning{EXO 0748-676: Search for pulsations }  
   \maketitle

\section{Introduction}
\label{sect:intro}

Gravitational Wave (GW) emission is an alluring phenomena, whose detection has eluded the physicists/astrophysicists so far. Direct detection of GWs is technically a very challenging task. Accreting neutron stars in Low Mass X-ray Binary (LMXB) systems, by virtue of their rapid oscillations, are potential sources for detection of GWs \citep{watts08}. Due to various reasons, such as, magnetic deformation \citep{cutler02, haskell08} and crustal mountains \citep{Bildsten98, Haskell06, Vigelius08}, these compact objects can develop a quadrupolar asymmetry which leads to GW emission. The GW emission is also believed to limit the spin frequency up to which the neutron stars can be spun-up by accretion \citep{Chakrabarty03}. In addition to being an important class of potential sources for GW detection, any detection of GWs from these sources may also provide important information about the structure and geometry of neutron stars. It can also be an important tool to understand the interaction of the neutron star with its accretion disk.  

Detection of GWs associated with spinning neutron stars will be an arduous job as it requires a huge computational power. The major hurdles include, low accretion rates and uncertainty in the measurement of spin and orbital parameters, of most of the astrophysical sources \citep{watts08, Watts09}. There are sources like Sco X-1, which are quite bright, but the spin and orbital parameters are poorly constrained. In case of sources emanating weak signals, the data ought to be folded in order to increase signal to noise ratio. But this folding requires a priori and well constrained ephemeris for spin and orbital parameters. If the pulse and orbit ephemeris is accurately known, it will require only a single trial to search for gravitational waves. However, when parameters are poorly constrained, a large number of trials are required and hence the searches become computationally untenable.

Among the galactic sources, the LMXBs harboring rapidly spinning neutron stars are the best targets for GW searches with LIGO and VIRGO \citep{Abadie10, Abbott10, Sengupta10}. Among all LMXBs, the spin and orbital parameters and their evolution are best constrained in the accreting millisecond pulsars. Hence, least amount of computational resources will be required for the search of gravitational waves. But the accreting millisecond pulsars are transient systems, active only during short outbursts and the long term mass accretion rate is not high enough for detectable GW emission. The magnitude of long term average flux of most of the accretion powered millisecond pulsars is of the order of 10$^{-11}$ erg cm$^{-2}$ s$^{-1}$, which is much smaller than that of Sco X-1 (of the order of 10$^{-7}$ erg cm$^{-2}$ s$^{-1}$).

Next best sources are the burst oscillation sources and the kHz QPO sources \citep{watts08, Watts09}. But the number of trial searches required in these sources can go up to 35 orders of magnitude, depending on the number of constrained parameters.

Among the persistent LMXBs, EXO 0748-676 is a promising source for detection of GWs. It has been studied extensively since its discovery in 1985 \citep{Parmar85}. Future monitoring of timing properties of this source, during the operation of Advanced LIGO \citep{Smith09, Harry10} will be possible with the large area X-ray detectors of $ASTROSAT$ \citep{Paul09}. Being an eclipsing source in a binary orbit with a period of 3.82 hr \citep{Parmar86}, the orbital period, period evolution timescale and mid-eclipse time are measured with good accuracy over a time base of more than 20 years \citep{Wolff09}. The source also shows irregular X-ray dips \citep{Parmar86, Wolff02}. The mass of the neutron star \citep{Cottam02, Pearson06, Ozel06, Munoz09} and its companion \citep{Parmar86} are also known with reasonable accuracy.

Since its discovery, EXO 0748-676 has been regularly monitored by various satellites. Most of the observations reveal a persistent luminosity of $\sim$ 10$^{36-37}$ (D/7.4 kpc)$^{2}$ erg s$^{-1}$. The long term 2-20 keV average flux is $\sim$ 2 $\times$ 10$^{-10}$ erg cm$^{-2}$ s$^{-1}$ \citep{Wolff08a}. It entered a quiescent phase in 2008, with a decline of up to 2 orders of magnitude in the flux level \citep{Wolff08a, Wolff08b, Hynes08}. The source has also repeatedly shown several Type-I X-ray bursts \citep{Gottwald86, Parmar86, Wolff02}. Burst oscillations at 44.7 Hz were discovered by \citet{Villarreal04}. They associated these oscillations with the spin frequency of the neutron star. However, \citet{Galloway10} recently found another millisecond oscillations feature at 552 Hz during some of the thermonuclear X-ray bursts. The actual spin period of the neutron star is not known.

In the present work, we have tried to verify whether the 44.7 Hz burst oscillations, seen during the thermonuclear bursts, can be associated with the spin frequency of the neutron star. A positive result would also allow measurement of the semi-amplitude of the binary orbit, and a monitoring of the spin and orbital parameters of this source would make deep search for GW
emission possible by allowing a long integration of GW data. 

\section{Observations and analysis}
\label{sect:Obs}

The data used in this work was obtained from observations made with the Proportional Counter Array (PCA: \citet{Jahoda96}) on board the Rossi X-Ray Timing Explorer ($RXTE$: \citet{Bradt93} ), on August 17, 1997 (Observation ID: 20082-01-02-00). The total exposure time was ~37ks, covering about 3 binary orbits. All the five PCUs were ON during the observation. The data was collected from the Science event mode (E$_{-}$125$\mu$s$_{-}$64M$_{-}$0$_{-}$1s), having a time resolution of 125 $\mu$s and 64 binned energy channels. The raw events were filtered using the \textsc{sefilter} and \textsc{seextrct} tool of the $ftools$ (version V6.5.1) package of the HEASOFT. The average source and background count rate during the observation was $\sim$155 and $\sim$45 counts/s, respectively, i.e. the source was about 4 times more intense as compared to the background. The light curve was barycentered using the $ftool$-\textsc{fxbary}. 

Figure 1 shows the barycentered light curve of EXO 0748-676, folded into 64 orbital phasebins with a period of 13766.78824 s (Wolff et al. 2009). The folded light curve was normalized by
dividing the count rate by the average source intensity. The folded light curve shows irregular dips along with a total eclipse lasting for about 0.04 orbital phase, i.e. 550 s.  

\begin{figure}
\centering
\includegraphics[height=3.3in, width=3.0in, angle=-90]{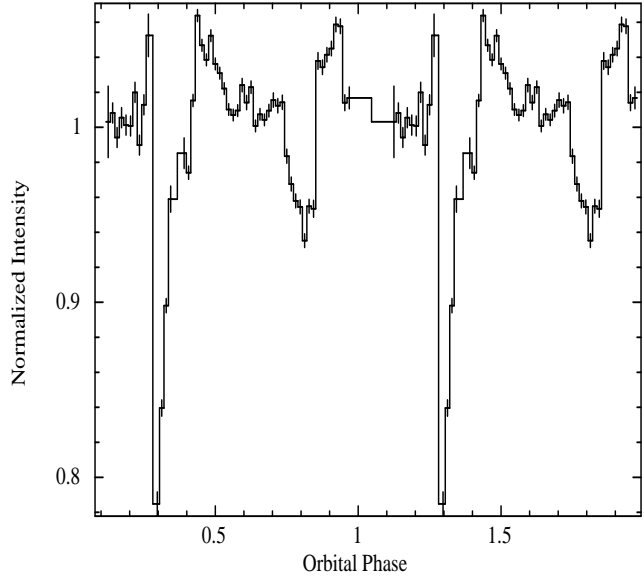}
\caption{Folded light curve of EXO 0748$-$676, obtained from observations made with $RXTE$-PCA. The light curve was folded into 64 phasebins with a period of 13766.78824 s. Two cycles
are displayed for clarity.}
\end{figure}

For a neutron star in a binary system, the arrival times of the pulses vary with the orbital motion of the pulsar. When the pulsar is at T$_{\pi/2}$ (the farthest end on the orbit from the
observer, also referred to as the superior conjunction), the photon takes a longer time (by ${a_{x} \sin i}\over{c}$) to arrive (here, $i$ is the source inclination angle and $a_{x}\sin i$ is the projected semi major axis). Therefore, assuming a circular orbit for this LMXB, the time delay of photon arrival time caused by the orbital motion can be expressed as:

\begin{eqnarray}
t_{arrival} = t_{emission} + \frac{a_{x} \sin i} {c} \cos \left( \frac{2 \pi
(t_{emission} - T_{\pi/2})} {P_{orbital}} \right)
\end{eqnarray}

where the $\rm{P}_{orbital}$ is the orbital period of the binary system.

In order to search for pulsations in EXO 0748-676, we first corrected the arrival time of the photons for different assumed values of a$_{x} \sin i$. 
Based on the known values of the mass function and the mass of the neutron star and its companion \citep{Munoz09}, the value of a$_{x} \sin i$ is expected to be around 730 lt-ms. The light curve was corrected for a$_{x} \sin i$ over a wide range of 100-1200 lt-ms, with a resolution of 2 lt-ms which is about 10\% of the pulse period around which the search was made.

After each correction of the light curve for the orbital motion, we searched for pulsations using the pulse folding and $\chi^{2}$ maximization $ftool$-\textsc{efsearch}. This is a well established technique and has been successfully used to determine the spin period and orbital parameters in various sources (Jain, Dutta \& Paul, 2007; Jain, Paul \& Dutta, 2010). The burst oscillations from EXO 0748-676 have been reported to be centered at 44.7$\pm$0.06 Hz (Villarreal $\&$ Strohmayer 2004). This indicates that the neutron star spin frequency probably lies between 0.02234 and 0.02240 s. Using the $ftool$-\textsc{efsearch}, the light curves were folded into 16 phasebins. We searched for 300000 trial periods, with a resolution of 1$\times$10$^{-9}$ s and centered at 0.02237 s. This enabled us to search for period over a range which is 5 times larger than the expected range for the neutron star spin period with a
maximum pulse phase smearing of 10\%. Each folded light curve was fitted with a constant and $\chi^{2}$ determined. If the trial period and the a$_{x} \sin i$ value are not correct, then the folded profile is smeared and the $\chi^{2}$ of the fit will be small, close to the number of phasebins. But if the trial period is correct, the pulse profile is reproduced correctly and a constant fit is expected to give a high $\chi^{2}$. Therefore, the trial period corresponding to a large $\chi^{2}$ will represent the true pulse period in the light curve, if any. In this process, the a$_{x} \sin i$ will also be determined.

Figure 2 shows a few samples of the $\chi^{2}$ distribution with the trial spin periods for different values of a$_{x} \sin i$. The value of a$_{x} \sin i$ used for orbital correction is mentioned in each case. We did not find a significantly high $\chi^{2}$ for any spin period in the entire range of a$_{x} \sin i$. This is shown in Figure 3, where a histogram has been plotted which shows the number of times a specific $\chi^{2}$ occurs for all the 300,000 trial spin periods. Histograms were created for the entire range of a$_{x} \sin i$ and were overlaid. A distribution centered at a $\chi^{2}$ of about 14-15 is seen in each case. A large deviation from this curve will indicate detection of a spin period for the neutron star. As seen in Figure 3, a $\chi^{2}$ of about 70 is seen in a few cases, which is an offset from the $\chi^{2}$ distribution curve. But detection significance of these deviations is poor as can be seen in Figure 2.

\begin{figure*}
\includegraphics[height=3.0in, width=3.0in, angle=-90]{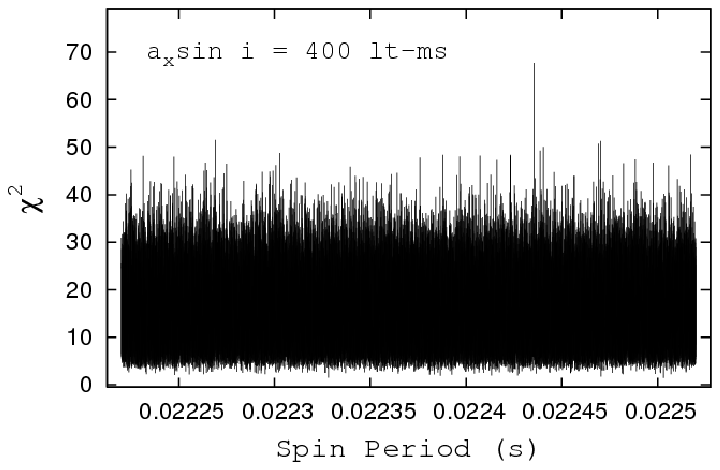}
\includegraphics[height=3.0in, width=3.0in, angle=-90]{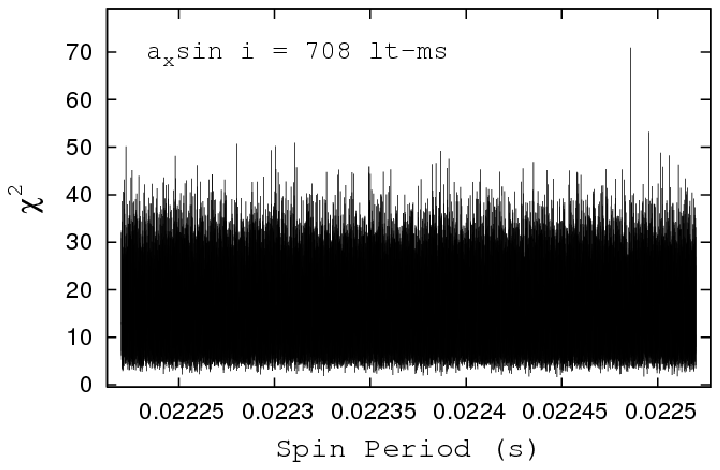}
 
 \includegraphics[height=3.0in, width=3.0in, angle=-90]{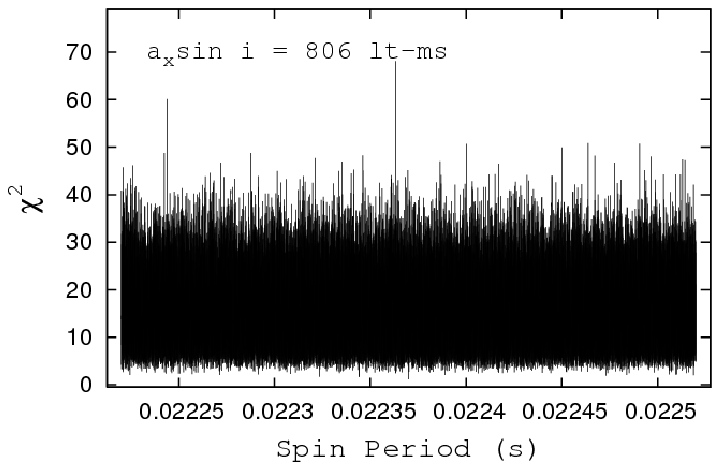}
 \includegraphics[height=3.0in, width=3.0in, angle=-90]{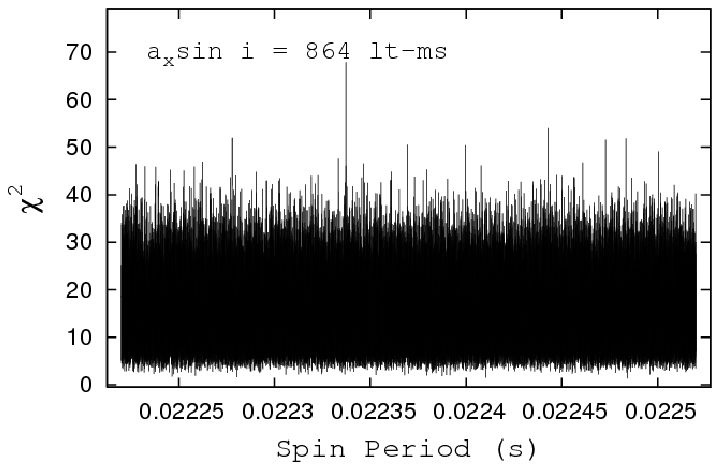}
\caption{A sample of $\chi^{2}$ variation with trial spin periods for different
values of a$_{x} \sin i$ for EXO 0748$-$676.}
\end{figure*}

\begin{figure}
\centering
\includegraphics[height=3.3in, width=3.0in, angle=-90]{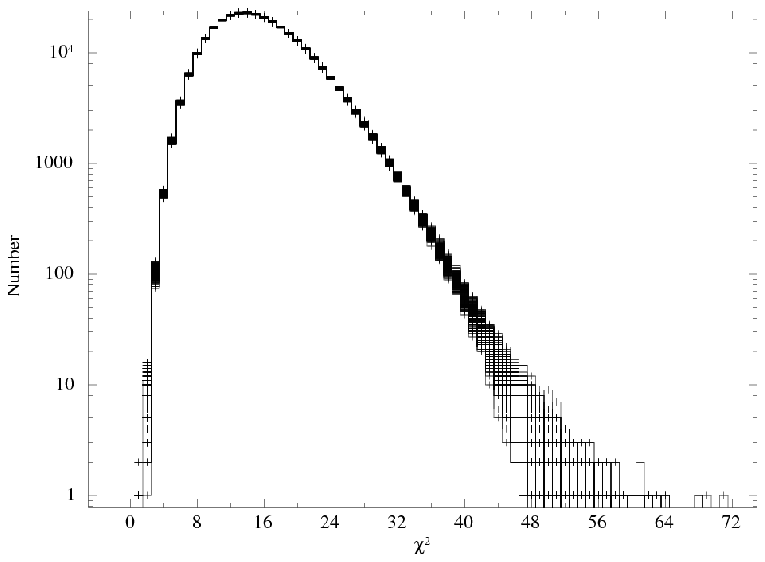}
\caption{Histogram showing the variation of $\chi^{2}$ for 300,000 trial spin periods of EXO 0748-676 and for the entire range of a$_{x} \sin i$.}
\end{figure}

In order to put an upper limit to the pulse amplitude present, if any, we corrected the light curve for the orbital motion with an arbitrary a$_{x} \sin i$. The light curve was then folded with a period of 0.00237 s. A model consisting of a sine curve was fit to the folded profile. This gave a 3 $\sigma$ upper limit of 0.47 \% for the pulse amplitude.

\section{Discussion}
\label{sect:discussion}

We have presented results from an X-ray timing analysis of the eclipsing LMXB, EXO 0748-676 in which we searched for pulsations at frequencies around the 44.7 Hz burst oscillations in
this source observed during some Type-I X-ray bursts (Villarreal $\&$ Strohmayer 2004). We conclude absence of any pulsations around the reported burst oscillation frequency. We have determined a 3$\sigma$ upper limit of 0.47 \% on the pulse amplitude.

Recently, Galloway et al. (2010) reported another burst oscillation feature at 552 Hz, which has higher detection significance and could be the true oscillation frequency. In order to
detect pulsations around 552 Hz during non-burst episodes, a similar analysis, as explained above is ongoing.

As mentioned before, detection of GWs is technically, an enormous task. The foremost sources for the detection of GWs are the neutron stars of/in LMXBs, where asymmetries are induced in the neutron star crust by accretion. The prerequisites for a potential target for GW searches, are, the strength of the signal and well constrained orbital and spin parameters. Sources like Sco X-1 and GX 17+2 may have a large GW amplitude, owing to their high X-ray flux and mass accretion rates ($\sim$ 3 $\times$ 10$^{-9}$ M$_{\odot}$ yr$^{-1}$  - 2 $\times$ 10$^{-8}$ M$_{\odot}$ yr$^{-1}$) \citep{Strohmayer05}. But the orbital and spin parameters are poorly constrained. The accretion powered millisecond pulsars, have well constrained orbital and spin parameters, but these are transient sources, and have low long-term average accretion rates (with magnitude less than 10$^{-11}$ M$_{\odot}$ yr$^{-1}$) \citep{Strohmayer05}.

EXO 0748-676 is definitely a promising source for detection of GWs. The long-term mass accretion rate of EXO 0748-676 is quite high ($\sim$2.9 $\times$ 10$^{-10}$ M$_{\odot}$ yr$^{-1}$ \citep{Wolff05}) and most of the orbital parameters are well constrained. If pulsations are detected, it will make this source an ideal candidate for search of GW emission. 

\normalem
\begin{acknowledgements}
This research has made use of $RXTE$ data obtained through the High Energy Astrophysics Science Archive Research Center (HEASARC) online service, provided by the NASA/ Goddard Space Flight Center. CJ acknowledges the financial support and hospitality at RRI during part of this work.
\end{acknowledgements}

\label{lastpage}
\end{document}